\documentclass[aps,prd,twocolumn,floatfix]{revtex4-2}

\usepackage{lipsum}
\usepackage{graphicx}  % needed for figures
\usepackage{dcolumn}   % needed for some tables
\usepackage{bm}        % for math
\usepackage{amssymb}   % for math
\usepackage{amsmath}
\usepackage{xcolor}
\usepackage{hyperref}
\usepackage{amsmath}
\usepackage{amsfonts}
\hypersetup{colorlinks=true, linkcolor=blue, citecolor=blue, urlcolor=blue}

\long\def\comment#1{}

\begin{document}

%\preprint{}

\title{\bf Search for galactic axions with a traveling wave parametric amplifier}

%authors and affiliations
%\comment{
\author{R.~Di Vora}
\author{A.~Lombardi}
\author{A. Ortolan}
\author{R.~Pengo} 
\author{ G. Ruoso} \email[Corresponding author: ]{Giuseppe.Ruoso@lnl.infn.it}
\affiliation{INFN, Laboratori Nazionali di Legnaro, 35020 Legnaro (Padova), Italy}

\author{C.~Braggio}
\affiliation{INFN, Sezione di Padova, 35100 Padova, Italy} \affiliation{Dipartimento di Fisica e Astronomia, 35100 Padova, Italy}
\author{G.~Carugno}
\author{L.~Taffarello}
\affiliation{INFN, Sezione di Padova, I-35100 Padova, Italy} 
\author{G. Cappelli}
\author{N.~Crescini}
\author{M. Esposito}\email[Present address: CNR-SPIN Complesso di Monte S. Angelo, 80126 Napoli, Italy]{}
\author{L. Planat}
\author{A. Ranadive} 
\author{N.~Roch}
\affiliation{Univ. Grenoble Alpes, CNRS, Grenoble INP, Institut N\'eel, 38000 Grenoble, France}

\author{D.~Alesini}  
\author{D.~Babusci}
\author{A.~D'Elia}
\author{D.~Di~Gioacchino}
\author{C.~Gatti} 
\author{C.~Ligi}
\author{G.~Maccarrone}
\author{A.~Rettaroli} 
\author{S.~Tocci} 
\affiliation{INFN, Laboratori Nazionali di Frascati, 00044 Frascati (Roma), Italy}
 
\author{D.~D'Agostino}
\author{U.~Gambardella}
\author{G.~Iannone}
 \affiliation{Dipartimento di Fisica E.R. Caianiello, 84084 Fisciano (Salerno), Italy} \affiliation{INFN, Sezione di Napoli, 80126 Napoli, Italy}

\author{P.~Falferi} 
\affiliation{Istituto di Fotonica e Nanotecnologie, CNR Fondazione Bruno Kessler, I-38123 Povo, Trento, Italy} \affiliation{INFN, TIFPA, 38123 Povo (Trento), Italy}

\collaboration{QUAX Collaboration}

\date{\today}

\begin{abstract}
A traveling wave parametric amplifier has been integrated in the haloscope of the QUAX experiment. A search for dark matter axions has been performed with a high Q dielectric cavity immersed in a 8 T magnetic field and read by a detection chain having a system noise temperature of about 2.1 K at the frequency of 10.353 GHz. Scanning has been conducted by varying the cavity frequency using sapphire rods immersed into the cavity. At multiple operating frequencies, the sensitivity of the instrument was at the level of viable axion models.

\end{abstract}

%\pacs{}
%\keywords{}

\maketitle


\section{\label{sec:intro}Introduction}

The axion is an hypothetical particle that arises from the spontaneous breaking of the Peccei-Quinn symmetry of quantum chromodynamics (QCD), introduced to solve the so-called strong CP problem~\cite{weinberg1978new,wilczek1978problem,peccei1977cp}. It is a pseudoscalar neutral particle having negligible interaction with the ordinary matter, making it a favourable candidate as a main component of dark matter~\cite{preskill1983cosmology}. Cosmological and astrophysical considerations suggest an axion  mass range $1~\mu\text{eV} < m_a < 10~\text{meV}$~\cite{irastorza2018new}. 

The hunt for axions is now world spread and most of the experiments involved in this search use detectors based on the haloscope design proposed by Sikivie~\cite{sikivie1983experimental,PhysRevD.32.2988}.
Among them are ADMX~\cite{braine2020extended,du2018search,boutan2018piezoelectrically,bartram2021search}, HAYSTAC ~\cite{backes2021quantum,zhong2018results}, ORGAN~\cite{mcallister2017organ,ORGAN2022}, CAPP-8T~\cite{choi2021capp,lee2020axion}, CAPP-9T~\cite{jeong2020search}, CAPP-PACE~\cite{kwon2021first}, CAPP-18T ~\cite{lee2022searching}, 
CAST - CAPP~\cite{Adair_2022}, CAPP-12TB~\cite{CAPP12TB},
GrAHal ~\cite{grenet2021grenoble}, RADES~\cite{melcon2020scalable,melcon2018axion,alvarez2021first}, TASEH~\cite{PhysRevD.106.052002}, QUAX~\cite{alesini2019galactic,alesini2021search,PhysRevD.106.052007,barbieri2017searching,crescini2018operation,crescini2020axion}, and KLASH~\cite{gatti2018klash,alesini2019klash}.
Dielectric and plasma haloscopes have also been proposed, the most notable examples being MADMAX~\cite{caldwell2017dielectric,Knirck_2021} and ALPHA ~\cite{lawson2019tunable,alpha2022,Alpha2023}, respectively.
The haloscope concept is based on the immersion of a resonant cavity in a strong magnetic field in order to stimulate the inverse Primakoff effect, converting an axion into an observable photon~\cite{al2017design}. To maximise the power of the converted axion, the cavity resonance frequency  has to be tuned to match the axion mass, while larger cavity quality factors $Q$ will result in larger signals.  Different solutions have been adopted to maximize the signal-to-noise ratio, facing the problem from different angles. Resonant cavities of superconductive and dielectric materials are becoming increasingly popular because of their high $Q$~\cite{di2019microwave,ahn2019maintaining,alesini2020high,alesini2021realization}.
In this work we describe the results achieved by the haloscope of the  QUAX--$a\gamma$ experiment, based on a high-$Q$ dielectric cavity immersed in a static magnetic field of 8 T.
Cooling of the cavity at $ \sim 100$  mK reduces thermal fluctuations and enables operation 
 of a traveling wave parametric amplifier (TWPA) at about the quantum limit.   
%The results obtained allow us to exclude values of $g_{a\gamma\gamma} > 0.729 \times10^{-13}$~GeV$^{-1}$ at 90\% confidence level (C.L.) in a region of mass, that is not currently accessible to other running experiments, centered at $42.8216\, \mu$eV and wide 1.32 neV. 
This is the first example of a high frequency haloscope (above 10 GHz) instrumented with a near quantum limited wide-band amplifier. A key step in the realization of an apparatus capable of searching for dark matter axions over an extended mass region. Section \ref{sec:setup} describes the experimental set-up with the characterization of all the relevant components, while in Section \ref{sec:results} details in the data analysis procedure are given. Since no dark matter signals have been detected, in the same Section  upper limits for the axion-photon coupling are deduced.

\section{\label{sec:setup}Experimental Setup}

\subsection{General description}

The core of the haloscope is a high-$Q$ microwave resonant cavity: at a temperature of about 4 K and under a magnetic field of  8\,T, we measured an internal quality factor of more than $9\times 10^6$\cite{PhysRevApplied.17.054013}.   It is based on a right-circular cylindrical copper cavity with hollow sapphire cylinders that confine higher-order modes around the cylinder axis.  The useful mode is a TM030 one, which has an effective volume for axion searches \cite{sikivie1983experimental,PhysRevD.32.2988} of $3.4 \times 10^{-2}$ liters at the resonant frequency of 10.353 GHz. Cavity tuning is obtained by displacing triplets of 2\,mm-diameter sapphire rods relative to the top and bottom cavity endcaps. Independent motion of the two triplets is obtained by  mechanical feed-throughs controlled by a room temperature motorised  linear positioner. 
%, where $C_{030}$ is a geometrical factor entering in the signal-power estimation in Equation \eqref{eq:power}. 
An extensive description of the cavity design and properties can be found in~\cite{PhysRevApplied.17.054013}.

%The cavity is hosted, together with some electronics, inside the Inner Vacuum Chamber (IVC) of a dilution unit, while the magnet are hosted inside a liquid-He cryostat at a temperature of about 4 K.

\begin{figure}[htb]
  \centering
      \includegraphics[width=0.4\textwidth]{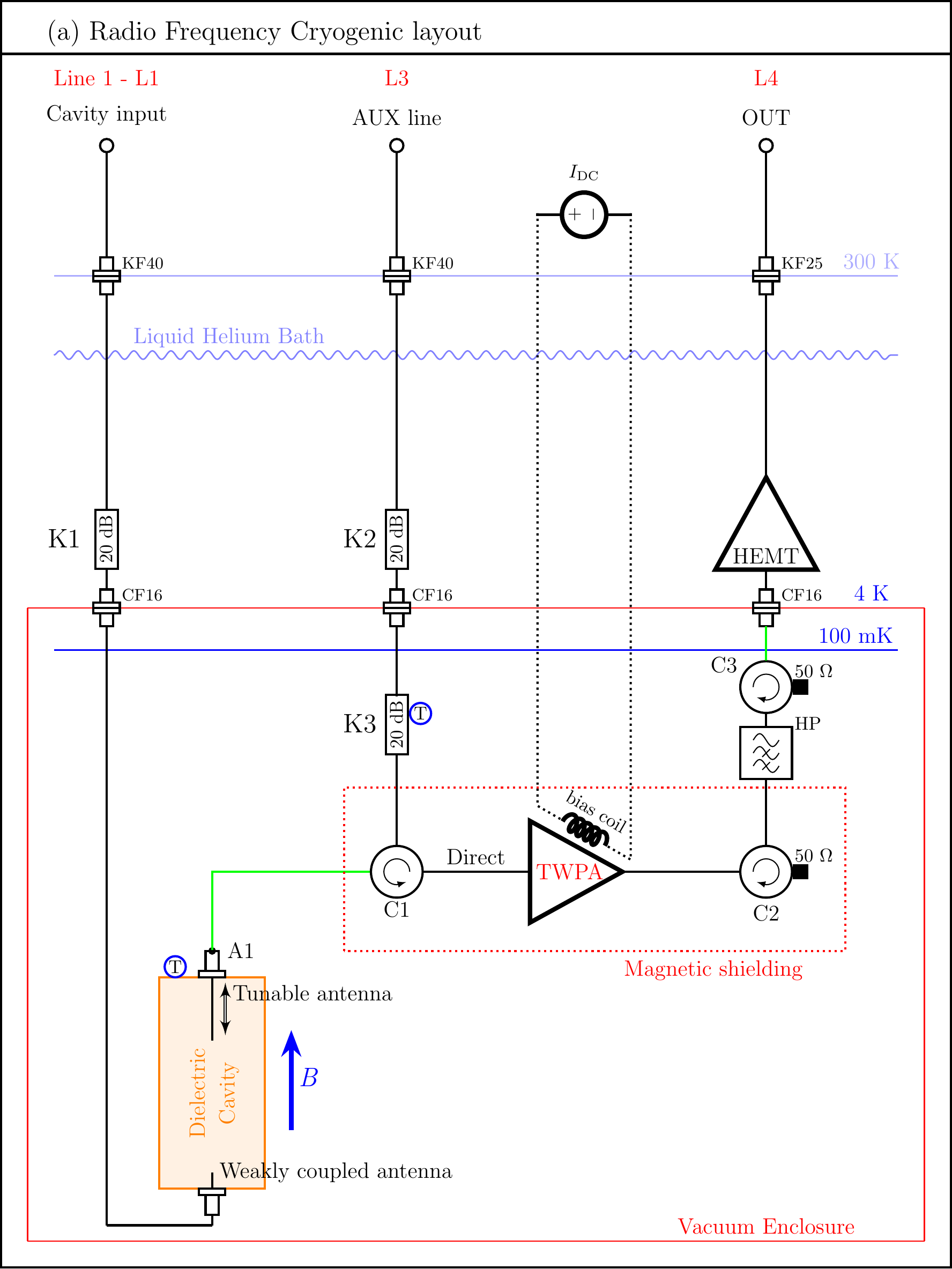}
\caption{\small Schematics of the experimental apparatus. The microwave cavity (orange) is immersed in the uniform magnetic field $B$ generated by the magnet. C1, C2 and C3 are circulators, HP is a 8 GHz high pass filter.
%HEMT is a High Electron Mobility Transistors. 
K1, K2, and K3 are attenuators, shown with their reduction factor in decibels. The components kept in the high vacuum region are enclosed in a red box. Circled T are RuO thermometers.  KF40 and KF25 are rf feed-through with ISO-KF 40 and ISO-KF 25 flanges, respectively. CF16 are rf feed-troughs with ConFlat 16 flange. Superconducting NbTi rf cables are in green. Details on the components are given in the Appendix.}
\label{fig:Apparatus}
\end{figure}

The layout of the measurement set-up is shown in Figure \ref{fig:Apparatus}. The microwave cavity is immersed in an 8 T  magnetic field generated by a
superconducting solenoid, and 
%150~mm diameter bore of  450~mm length superconducting magnet.  
it is read by a tunable dipole antenna with coupling $\beta$. 
The antenna is the inner core of a superconducting rf cable, for which the final dielectric insulation and metallic shielding have been removed for a length of 12 mm. The antenna position is determined by a spring placed between the cavity top and  the outside of the rf cable and acting against an electrical  motorised linear drive placed at room temperature and connected with a thin steel wire. Precise positioning with an electronic controller is possible over a length of about 20 mm,  that allows for $\beta$ values in the range 2 to 20. Due to tight mechanical constraints, the cavity works only in the overcoupled regime. A weakly coupled port (coupling about 0.01) is used for calibration purposes and is connected to the room temperature electronics by means of line L1. Cabling losses and the use of the attenuator K1 (20 dB) ensure an attenuation of about 30 dB for thermal power inputs from room temperature.

The tunable antenna output is fed onto a circulator (C1)  using a superconducting NbTi cable. C1 is  directly connected to the input of a  TWPA\cite{Ranadive2022}, which serves as pre-amplifier of the system detection chain. Further amplification at a cryogenic stage is done using a low-noise high-electron-mobility transistor (HEMT) amplifier.
%(Low Noise Factory model LNA4-16). 
In order to avoid back-action noise from the HEMT, a pair of isolators (C2 and C3) and an 8 GHz high-pass filter are inserted between the TWPA and the HEMT. The output of the cryogenic HEMT is then delivered to the room temperature electronics using line L4.
The auxiliary line L3 is used for calibrations and to provide a pump signal to the TWPA: it is connected to the line L4 by means of the circulator C1, and 20+20 dB of attenuation prevents thermal leakage from room temperature components and from the bath at 4 K. 
Finally, a dc current source is connected to a superconducting coil used to bias the TWPA.
Following  Figure \ref{fig:Apparatus}, all components
within the red box "Vacuum enclosure" are thermally anchored at the 
mixing chamber of a dilution unit 
%50 mK
%below the horizontal blue line sectioning the 4K region 
and enclosed in a vacuum chamber immersed in a liquid helium dewar. Two Ruthenium Oxide thermometers  measures the temperature of the cavity and of attenuator K3, respectively. 

\begin{figure}[htb]
  \centering
            \includegraphics[width=0.35\textwidth]{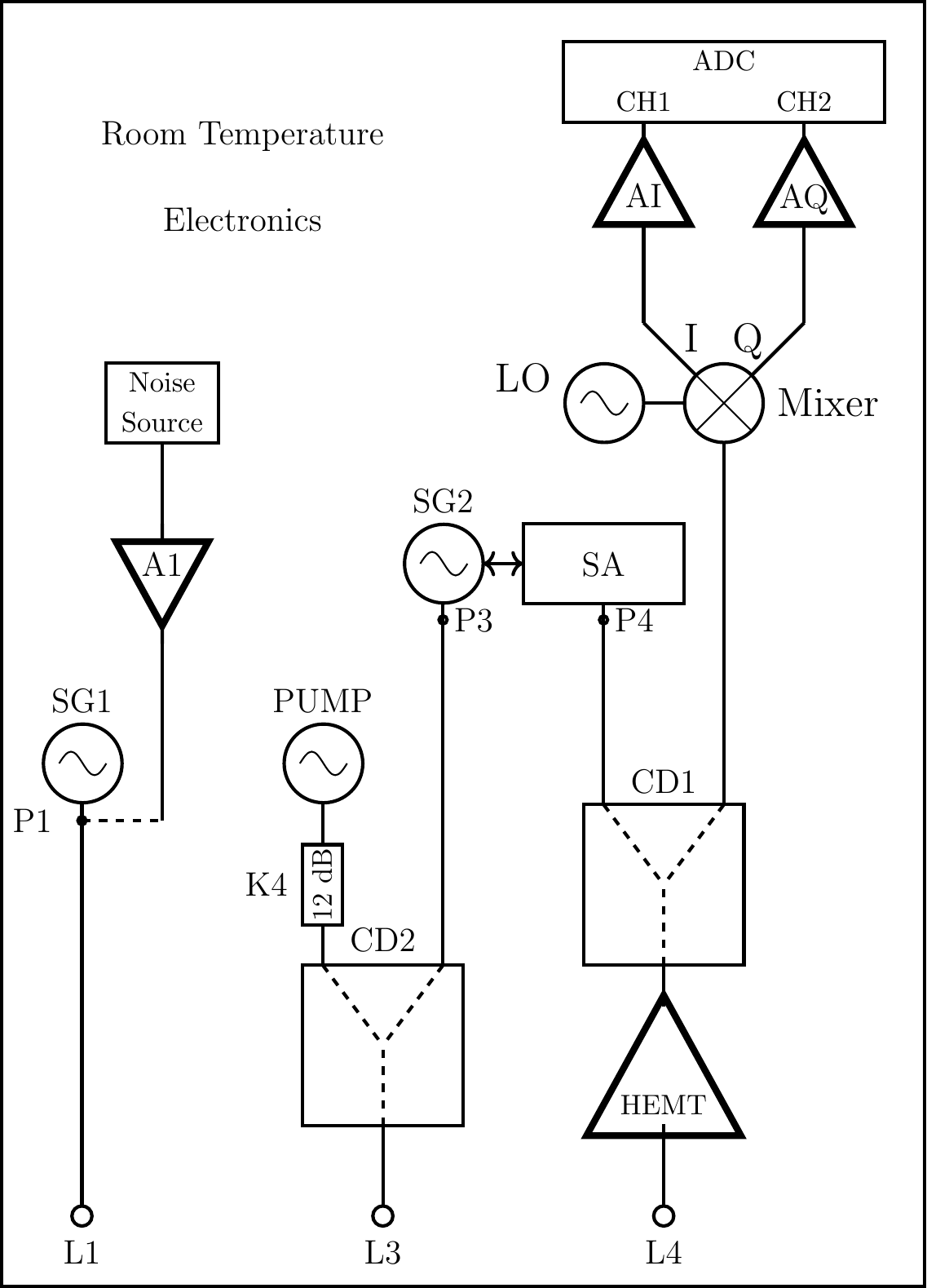}
\caption{\small Room temperature electronics. CD1 and CD2 are power splitters/combiners; LO, SG1, SG2 and PUMP are signal generators; 
%HEMT is  a High Electron Mobility Transistor; 
SA is a spectrum analyser; AI, AQ are low frequency amplifiers; A1 is a rf amplifier; ADC is an analog to digital converter with inputs CH1 and CH2. The signal generator SG2 can be controlled by the spectrum analyser for tracking to measure the cavity reflection spectrum. The lines L1, L3, and L4 are connected to the corresponding lines in Figure \ref{fig:Apparatus}.}
\label{fig:Apparatus2}
\end{figure}

The room temperature electronics scheme is shown in Figure \ref{fig:Apparatus2}. Signals from line L4 are amplified by a second HEMT and split by a power divider. 
One of the divider two outputs is fed into a spectrum analyser (SA), used for diagnostic and calibration. The input of the SA is referred as measurement point P4. The other output of the divider is down-converted using a mixer 
%(MITEQ IRM0812LC2Q)  
driven by the signal generator LO,
delivering a power of 12 dBm at a 
%(Keysight N5183B), 
 frequency about 500 kHz below the cavity resonance. 
 %The mixer input level for the local oscillator input has been set to 12 dBm.
 The room temperature chain is the same used in our previous measurements \cite{alesini2021search}:  the low frequency in-phase and quadrature outputs of the mixer are amplified 
 %(FEMTO DHPVA-100) 
 and then sampled with a 2 Ms/s analog to digital converter (ADC) and stored on a computer for off-line data analysis.
Data storage is done with blocks of about 4 s of sampled data for both output channels of the mixer.

By using the external source control option of the spectrum analyser SA it is possible to use the generator SG2 as a tracker to measure the cavity reflection spectrum 
%The room temperature electronics feature also a Vector Network Analyser (VNA) for measurement of 
S43 (input from line L3 - output from line L4). The reflection spectrum provides information on the loaded quality factor $Q_L$, resonance frequency $f_c$ and coupling $\beta$ of the tunable antenna. A diode noise source, having an equivalent noise temperature of about $10^4$ K,  can be fed to line L1 for testing after being amplified in such a way to have an equivalent noise temperature inside the microwave cavity slightly in excess of the thermodynamic temperature. A microwave signal generator and a microwave spectrum analyser are  used for the measurement of the  system noise temperature as described below. All rf generators and the spectrum analyser are frequency locked to a Global Positioning System disciplined reference oscillator.

% end changed text

\subsubsection{Cryogenic and vacuum system} 
The cryogenic and vacuum system is composed of a dewar and a $^3$He–$^4$He  wet dilution refrigerator. 
The dewar is a cylindrical vessel of height 2300 mm, outer diameter 800 mm and inner diameter 500 mm.
%(made by Precision Cryogenics System Inc). 
The dilution refrigerator is a refurbished unit 
%(made by Leiden Cryogenics Inc.)
previously installed in the gravitational wave bar antenna Auriga test facility~\cite{Marin_2002}. Such dilution unit (DU) has a base temperature of 50 mK and cooling power of 1 mW at 120 mK.
The DU is decoupled from the gas handling system through a large concrete block sitting on the laboratory ground via a Sylomer carpet where the Still pumping line is passing. This assembly minimizes the acoustic vibration  induced on the TWPA,  which is  rigidly connected to the mixing chamber. 
%A picture of the setup with the open cryostat is shown in figure ??. 
Once the Helium dewar has been filled up with liquid helium the DU column undergo a fast pre-cooling down to liquid-helium temperatures via helium gas exchange on the Inner Vacuum Chamber (IVC). This cooling-down operation takes almost 4 hours. When a temperature of 4 K has been reached, the pre-cooling phase ends, the inner space of the IVC is evacuated.  From that point on the dilution refrigerator takes over and the final cooling temperature slightly above 50 mK is attained after about 5 hours. 
%No charcoal pump was present in the dilution unit cooled system. 
A pressure of around 10$^{-7}$ mbar was monitored without pumping on the IVC room temperature side throughout all the experimental run.
Temperatures are measured with a set of different thermometers. Most of them are used to monitor the behaviour of the dilution unit. 
%Two RuO thermometers are used to monitor the temperature of the microwave cavity and of the attenuator K3.%It is to be noted that the 20 dB attenuator close to C1 is thermally anchored to the mixing chamber. 

\subsubsection{Magnet system}

A NbTi superconducting solenoid provides the background field of 8 T (value at centre of the magnet), charged at a final current of 92 A with a ramp rate not exceeding 7 mA/s to reduce eddy currents losses in the ultra-cryogenic  environment.   When the current reaches the nominal value, a superconducting switch can be activated and the magnet enters the persistent mode.  In this mode, the stability assures a loss of the magnetic field lower than 10 ppm/h. For this measurement campaign such mode was actually not used, and the magnet was kept connected to the current source. The magnet has an inner bore of diameter 150 mm, and a length of 450 mm. When the magnet is driven by the 92 A current, the effective squared field  over the cavity length amounts to 50.8 T$^2$.

If the magnet was not shielded,  the TWPA and sensible electronics would be exposed to a stray field in the range of $0.2-0.3$ T along its length, well above the operative conditions. In order to reduce the stray field in the area of the TWPA, a counter field is locally generated by an additional superconducting solenoid, made of NbTi superconducting wire (0.8 mm diameter) wound on an Aluminium reel. The inner diameter of this winding is 77.8 mm and its height is 250 mm. The counter field winding is biased in series to the main 8 T magnet, so it is able to reduce the field in the volume occupied by the TWPA, at any field strength, to a mean value of 0.04 T.
%The remaining field is shielded by a lead envelope, at 100 mk, surrounding the JPA which has a critical field of 800 Gauss at this temperature.
To shield such remaining field a hybrid box  encapsulates the two circulators C1 and C2 and the TWPA. This hybrid box is constituted by an external box of lead and an internal one of CRYOPERM\textcircled{R}. The box dimension is $35\times 65\times 210$ mm$^3$, with one small base opened to allow cabling, and is thermally anchored to the DU mixing chamber.

\begin{figure}[htb]
  \centering
      \includegraphics[width=0.45\textwidth]{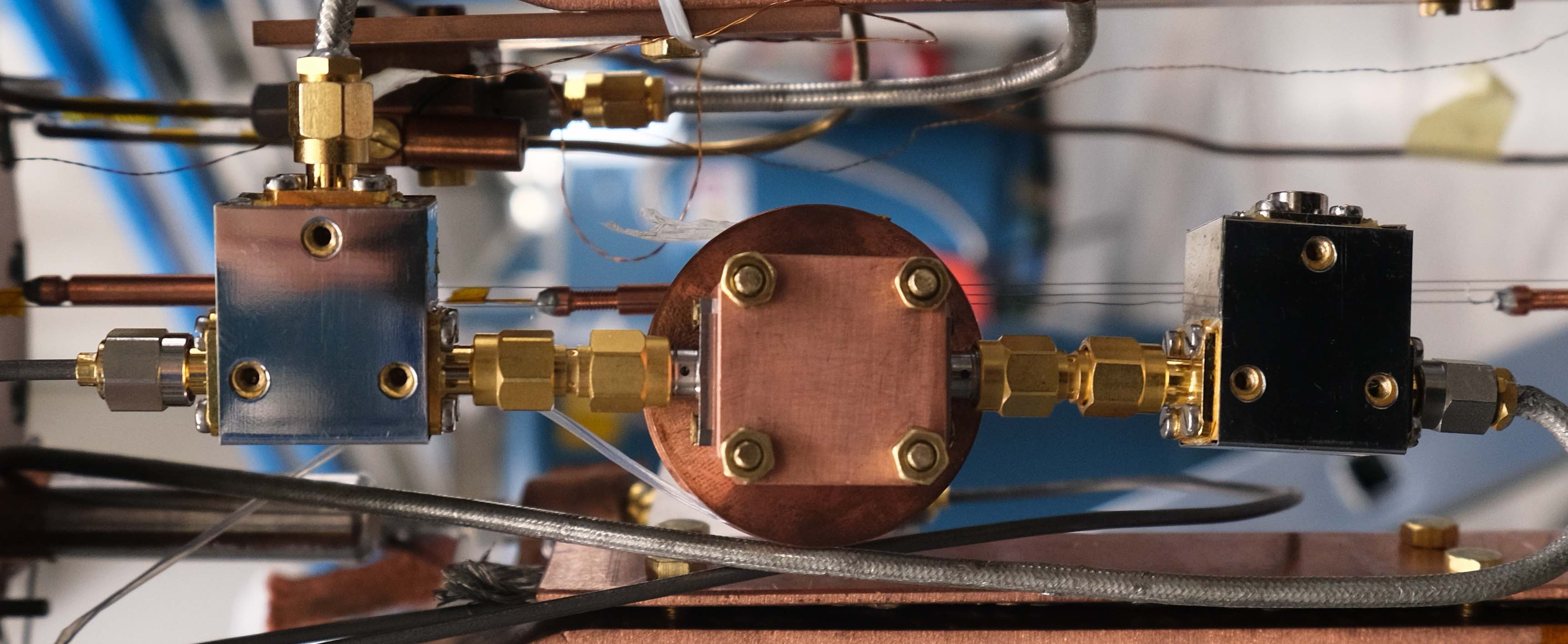}
\caption{\small Picture of the traveling wave parametric amplifier. The TWPA is enclosed in the central square box, the circle on the background is the holder of the superconducting coil providing the bias magnetic field. The two 3-port boxes are isolator C2 (on the right) and circulator C1 (on the left).  }
\label{fig:TWPA}
\end{figure}

\subsubsection{Amplifier characterization}
\begin{figure}[htb]
  \centering
        \includegraphics[width=0.45\textwidth]{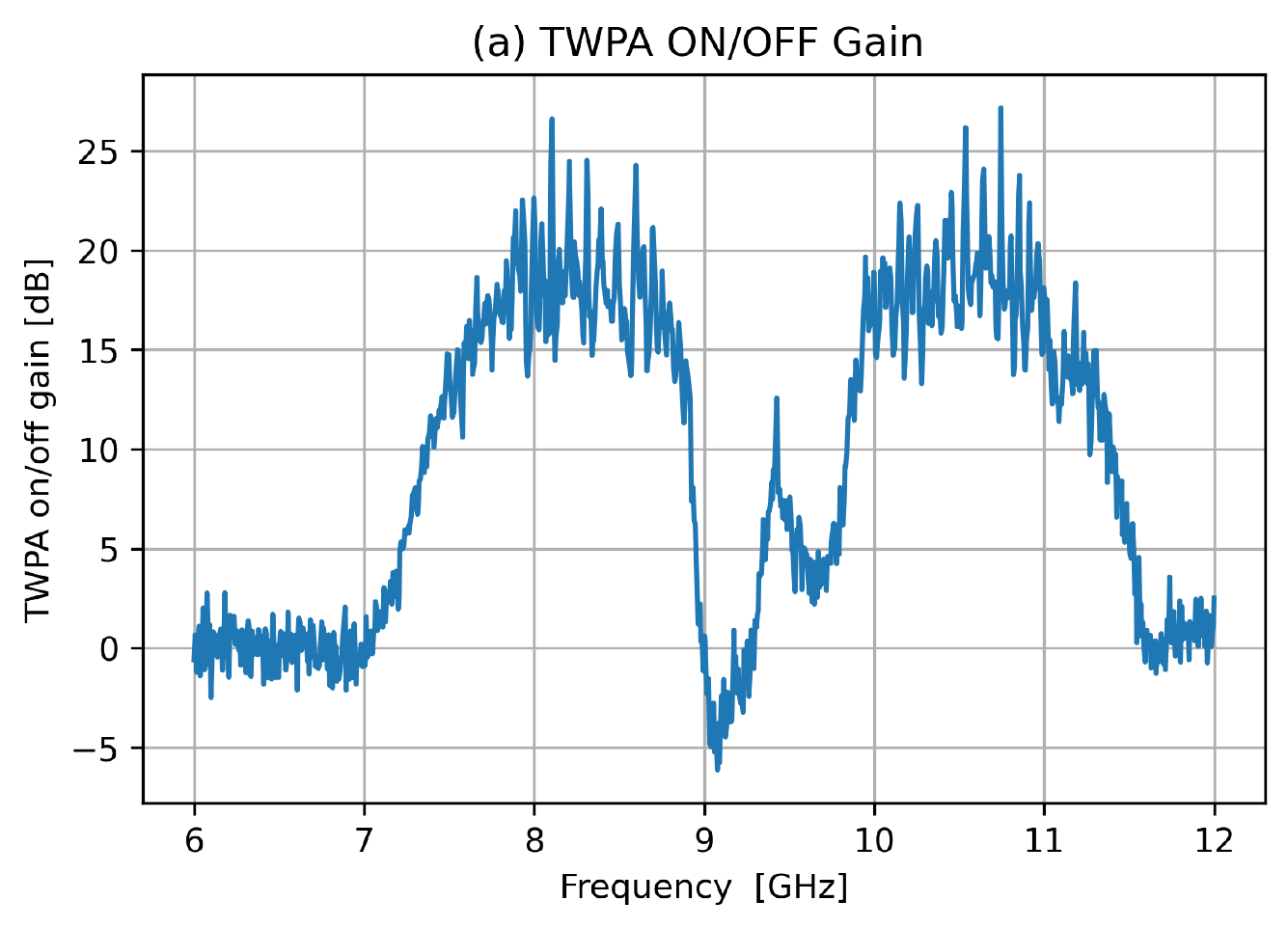}
              \includegraphics[width=0.45\textwidth]{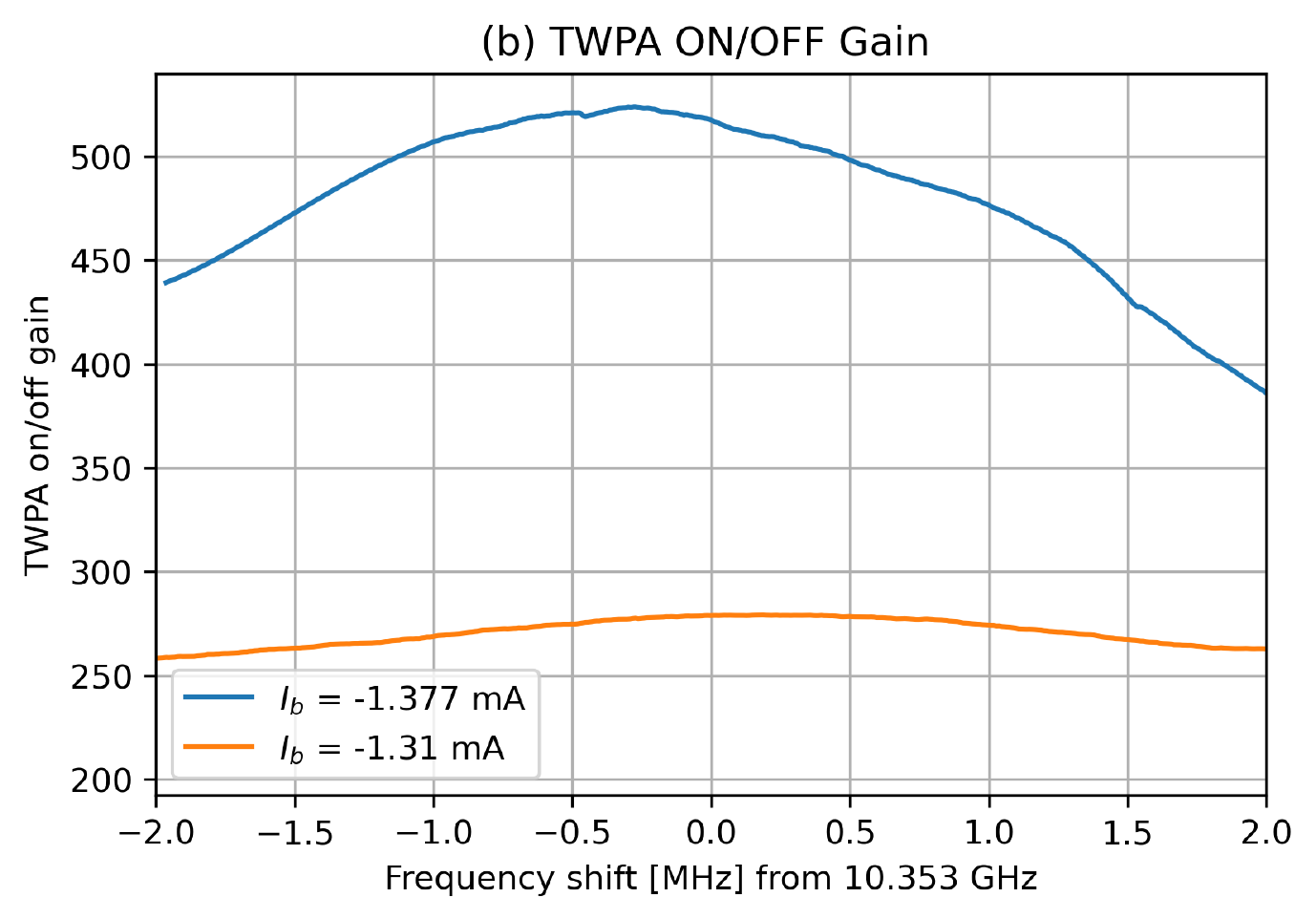}
\caption{\small Wide (a) and small (b)  bandwidth gain of the traveling wave parametric amplifier. The gain has been estimated by the changes in transmission between the two cases PUMP ON and PUMP OFF \cite{Braggio_2022}. (b) the two curves have been obtained with two different values of bias $I_b$ on the TWPA.  }
\label{fig:gain}
\end{figure}

The TWPA, see Figure \ref{fig:TWPA}, has been characterised following the procedure described in Ref. \cite{Braggio_2022}. In particular its working point in terms of bias current $I_b$,  frequency $f_P$ and  power $A_P$ of the PUMP generator of Figure \ref{fig:Apparatus2}, has been chosen in order to minimize the system noise temperature $T_{\rm sys}$ at the cavity unshifted frequency $\nu_c$. The performances of the TWPA have been measured several times with the magnetic field off, and then with the magnet energized once to 4 T and once to 8 T. All the resulting values of $T_{\rm sys}$ are compatible, around 2.0 K.
During the magnet current ramp up we monitored the wide bandwidth gain of the amplifier, to look for possible variations of the working point induced by the stray field passing through the shielding. Since no changes have been observed, one can conclude that the residual field is much below one flux quanta.
The wide bandwidth gain of the amplifier is shown in Figure \ref{fig:gain}(a). The PUMP frequency is set to $f_P=9.4181$ GHz, with $A_P=-16.5$ dBm, 
%(NOTA Attenuatore 12 dB), 
and $I_b=-1.38$ mA. It is evident from the figure that large (10 dB) oscillations of the gain are present at frequencies corresponding to higher gain.
%useful region, i.e. in the 10 - 11 GHz interval. 
By precisely varying bias and PUMP frequency it is possible to align a gain maximum to the cavity frequency: a gain maximum is normally equivalent to a minimum system noise temperature. Figure  \ref{fig:gain}(b) shows the gain in a 4 MHz interval centred at the cavity resonant frequency. The two gain profiles are obtained with two different values of bias $I_b$: different values of bias set different working points for the TWPA, with corresponding different gain value and profile. In general higher gains means a much sharper gain profile, but even for the sharpest one a useful region of flat gain of about 1 MHz is obtained.

% Requires the booktabs if the memoir class is not being used
\begin{table}[htbp]
   \centering
   %\topcaption{Table captions are better up top} % requires the topcapt package
   \begin{tabular}{|c|c|c|c|c|c|} % Column formatting, @{} suppresses leading/trailing space
\hline
n & Magnetic  & Cavity  & $T_{\rm sys}$ (K) & K3 & $T_{\rm sys}$ (K)  \\
 & field (T) & Temp. (K) &  On  Res. & Temp. (K) & Off Res.\\

\hline
1 & 0 & 0.12 & 2.12 $\pm$ 0.05 & 0.18 & $2.22 \pm 0.06$  \\
\hline
2 & 0 & 0.12 & 2.04 $\pm$ 0.03 & 0.19 & $1.94 \pm 0.03$\\
\hline
3 & 4.0 & 0.13 & 2.11 $\pm$ 0.03 & 0.22 & $2.16 \pm 0.03$ \\
\hline
4 & 0 & 0.12 & 1.89 $\pm$ 0.04& 0.18 & $1.98 \pm 0.05$ \\
\hline
5 & 8.0 & 0.11 & 2.23 $\pm$ 0.06 & 0.18 & $2.26 \pm 0.06$\\
\hline
   \end{tabular}
   \caption{Summary of measurement of system noise temperature $T_{\rm sys}$ at the cavity main frequency $\nu_c \simeq10.353$ GHz. "Res." stands for Resonance and "Temp." for Temperature. }
   \label{tab:tsys}
\end{table}

Table \ref{tab:tsys} shows the measured values of system noise temperatures, all of them measured at the test frequency  $\nu_c=10.353$ GHz.
%with the pump at $f_P=9.4181$ GHz. 
The table shows cavity  and attenuator temperatures, which contribute in different ways to the noise: the off cavity resonance value refers to a frequency 1 MHz detuned by $\nu_c$, where only the attenuator noise is seen by the TWPA, while on resonance a combination between cavity noise and attenuator noise forms the input noise. Only for the case of critical coupling ($\beta =1$), the noise at the cavity frequency is determined only by the cavity temperature. Except for the case n=5 (Magnetic field 8 T), having $\beta\simeq22$, the other measurements have $\beta\simeq 3$.   One extra measurement has been done at the frequency $\nu_{c,2} = 9.404 061$ GHz, where another cavity mode is present. For such measurement the PUMP frequency was set to $f_{P,2}=8$ GHz, for a  resulting $T_{\rm sys}(\nu_{c,2})=2.1\pm0.1$ K. 

For the selected cavity mode TM030,
%and in the absence of the external field,
the average value is $T_{\rm sys}^{\rm avg}=2.06 \pm 0.13$ K on resonance, and $T_{\rm sys}^{\rm avg}=2.07 \pm 0.14$ K off resonance. The central values come from the weighted average, while their errors are the standard deviation of all the values, showing a much wider distribution compared with the error of the single values.
%In the presence of the external magnetic field a slight degradation seems to be present, which could be due for example to an increase loss in the superconducting cable connecting the cavity main antenna to the circulator C1.
The resulting gain for the detection chain, from point A1 in Figure \ref{fig:Apparatus} to point P4 in Figure \ref{fig:Apparatus2} is $74.7\pm0.1$ dB. We have also obtained the gains for the other two rf lines, resulting in $-61.3\pm 0.1$ dB from point P1 to A1, and $-50.9\pm 0.1$ dB from point P3 to A1. All these gain values are those in the presence of magnetic field at 8 T.

We can try now to evaluate all the various contributions to the measured noise level.
In the quantum regime ($k_B T \ll h\nu$), the number of noise photons at frequency $\nu$ emitted by a thermal source is given by 
\begin{equation}
N\left(\nu,T\right) = \frac{1}{2}\coth\left( \frac{h \nu}{2k_B T} \right) , 
\end{equation}
where $h$ is Planck constant, $k_B$ is Boltzmann constant, $T$ is a thermodynamic temperature.
At the considered temperatures, the noise is entirely due to quantum fluctuations, as the contribution of the thermal photons is negligible.

At a given signal frequency $\nu_s$, the noise power spectral density at the output of the HEMT amplifier is  
\begin{align}
\begin{split}
\mathrm{PSD}_\mathrm{HEMT}(\nu_s) = & G_\mathrm{HEMT} \big[ N_\mathrm{HEMT} + \left( 1-\Lambda_2 \right)N_2 +  \\
							  & \Lambda_2 G_\mathrm{TWPA} \big( N_\mathrm{TWPA} + \left( 1-\Lambda_1 \right)N_1 + \\ 
							  &  \Lambda_1N(\nu_s,T_s) + \Lambda_1N(\nu_i, T_i) \big) \big] h\nu_s
\label{eq:psd_all}
\end{split}
\end{align}
where $G_\mathrm{TWPA}$ is the net gain of the TWPA, $N_\mathrm{TWPA}$ and $N_\mathrm{HEMT}$  are the added noise of the TWPA and HEMT, respectively. $\Lambda_1<1$ is the transmission  of the lossy chain from point A1 to the TWPA and, analogously, $\Lambda_2 <1 $ is the transmission from the output of the TWPA to the HEMT; $N_1$ and $N_2$ are the noise contributions coming from a simple beam splitter model for a lossy element.

$N(\nu_s,T_s)$ and $N(\nu_i,T_i)$ represent the quantum noise contributions
at the signal frequency equal to the cavity frequency, $\nu_s = \nu_c$, and at the idler frequency,  $\nu_i = 2 f_P - \nu_c$, respectively\cite{Braggio_2022}.
At the idler frequency the noise source is the attenuator K3, whose temperature is measured by one of the two thermometers. Its temperature has actually to be increased by the power leakage coming from the 4 K and room temperature stages. Considering the attenuation of K2 and K3 and cabling losses, we added 40 mK to the thermodynamic temperature of K3. At the signal frequency, the effective temperature is an intermediate value between the cavity temperature and the temperature of K3, the exact value depending on the coupling $\beta$: for $\beta=1$ we have just the cavity temperature.

The line transmissions are estimated at room temperature for the non-superconducting cabling, resulting in $\Lambda_1\simeq - 0.3\,\mathrm{dB}$ and $\Lambda_2\simeq - 0.7\,\mathrm{dB}$, which in linear units are close to 1. Such transmissions show low losses for the lines and allowed us to neglect the noise contributions $N_1$ and $N_2$ in Equation (\ref{eq:psd_all}). 
In addition, the high gain of the HEMT allows us to neglect all the noise contribution entering after its amplification.

With such simplifications, Eq.\,(\ref{eq:psd_all}) can be recast to estimate the total system noise (referred at the point $A1$), 
\begin{equation}
\mathrm{PSD}_\mathrm{A1} (\nu_s) = \frac{\mathrm{PSD}_\mathrm{HEMT}(\nu_s)}{\Lambda_1\Lambda_2 G_\mathrm{TWPA}G_\mathrm{HEMT}} = N_{\rm sys} h \nu_s, 
\end{equation}
where we have defined $N_{\rm sys}$ as the total number of  noise photons for the system,  obtaining
\begin{equation}
 N_{\rm sys} \simeq  N(\nu_s,T_s) + N(\nu_i,T_i) + \frac{N_\mathrm{TWPA}}{\Lambda_1} + \frac{N_\mathrm{HEMT}}{\Lambda_1 \Lambda_2 G_\mathrm{TWPA}} .
\label{eq:psd_a1}
\end{equation}
Giving the measured value $T_{\rm sys}= 2.06 \pm 0.13 $ K, we obtain in terms of photons a noise level:
\begin{equation}
 N_{\rm sys}  = \frac{k_B}{h \nu_s} T_{\rm sys}= 4.2 \pm 0.3.
\end{equation}

% Requires the booktabs if the memoir class is not being used
\begin{table}[htbp]
   \centering
   \begin{tabular}{|c|c|c|} % Column formatting, @{} suppresses leading/trailing space
   \hline
   Term & Value (K) & $N$ photons \\
   \hline
  $ N(\nu_s,T_s) $ & 0.27 & 0.5 \\
    \hline
  $ N(\nu_i,T_i) $ & 0.27 & 0.7 \\
  \hline
  ${N_\mathrm{HEMT}}/{\Lambda_1 \Lambda_2 G_\mathrm{TWPA}} $ & 0.39 & 0.8 \\
  \hline
  \hline
  $ N_{\rm sys}$ & 2.06 & 4.2 \\
  \hline
  \hline
  $ {N_\mathrm{TWPA}}/{\Lambda_1} $ & & 2.2 \\
  \hline

   \end{tabular}
   \caption{Noise contributions to the system noise. In the column Value, the noise is expressed in terms of Equivalent Noise Temperature. The following values have been used: $\nu_s=$ 10.353 GHz, $\nu_i=$ 7.94 GHz, and the thermodynamic temperatures $T_s= 0.16$ K, $T_i= 0.22$ K. $ {N_\mathrm{TWPA}}/{\Lambda_1} $ is the difference between the measured $  N_{\rm sys}$ and the sum of the preceeding terms.}
   \label{tab:noise}
\end{table}

In order to disentangle the contribution of the HEMT, we measured the system noise with the TWPA off (with PUMP OFF and without bias). The resulting noise temperature in this case was $49\pm1$ K, with a total gain reduced by a factor 125. The contribution of the HEMT to the total noise temperature is then $0.39 \pm 0.01$ K. Table \ref{tab:noise}
summarises the noise contributions and allows to derive the TWPA added noise as $N_\mathrm{TWPA}  \simeq $ 2.1 photons at the frequency of 10.353 GHz.

\subsubsection{Data taking}\label{sec:datataking}

\begin{table}[htbp]
   \centering
   \begin{tabular}{|c|c|c|c|c|} % Column formatting, @{} suppresses leading/trailing space
   \hline
   RUN & $\nu_c$ & Duration & $T_c$ & $T_{\rm K3}$ \\
   n & (GHz)  & (s) &(mK)& (mK) \\
   \hline
   389 & 10.353 525 & 2000 & 113 & 177 \\
   \hline
   392 & 10.353 499  & 2000 & 111 & 178\\
   \hline 
   394 & 10.353 473 & 2000 & 112 & 181 \\
   \hline
   395 & 10.353 473 & 4000 &113 & 185\\
   \hline
   397 & 10.353 444 & 2000 & 114 & 182 \\
   \hline
   399 & 10.353 424 & 4800 & 112 & 177 \\
   \hline
   401 & 10.353 399 & 28000 & 110 & 176 \\
   \hline
   404 & 10.353 368 & 2000& 110 & 176 \\
   \hline
   \end{tabular}
   \caption{Summary of the RUNs performed for the axion dark matter search. The cavity frequency is the value determined via a fast tracking spectrum on the SA. The frequency of the LO has been set to 10.353 GHz for all RUNs but RUN 404, where it is set to 10.3529 GHz. $T_c$ and $T_{\rm K3}$ are the cavity and attenuator K3 temperatures, respectively.}
   \label{tab:runset}
\end{table}

The search for axion dark matter has been performed over a time span of about 17 consecutive hours. The cavity antenna coupling has been set to overcritical, with the target to have a loaded quality factor about 4 times smaller than the axion one\cite{turner1990periodic}. This is important for what concern data analysis. Data taking is divided in different units that we usually call RUN, each RUN differing from another normally for the cavity central frequency that can be varied with the sapphire triplets described above. The detection chain system noise temperature and gain have been measured once at the beginning of the scanning session, and we monitored the time stability of the gain by injecting a rf pure tone of amplitude -90 dBm on line L3 using SG2 at a frequency 900 kHz above the LO frequency. The LO frequency is chosen in order to keep the cavity frequency in a central band of the ADC working region.
Each data taking step is composed of the following actions:
\begin{description}
\item[action 1] Set the cavity frequency to the desired value by moving the sapphire triplets. Set the LO frequency: this is actually not done for every step, since normally the change in cavity frequency is much smaller of the ADC bandwidth.
\item[action 2] Measure a cavity reflection spectrum by tracking generator SG2 with the spectrum analyser SA. Data are saved on a file to be fitted to deduce the antenna coupling $\beta$.
\item[action 3] Measure a cavity transmission spectrum by using the noise source on line L1 and acquiring data with the ADC. Data are also taken with the the spectrum analyser SA for quick analysis. The number of 4 s blocks acquired with the ADC is normally about 30.
\item[action 4] Measure the cavity output with all the inputs off, acquiring data with the ADC. This is the  axion search data, and normally we collect 500 blocks of 4 s length each. Some RUNs have been done with a larger number of blocks. Again, data are also taken with the the spectrum analyser SA for quick analysis.
\end{description}

\begin{figure}[htb]
  \centering
      \includegraphics[width=0.45\textwidth]{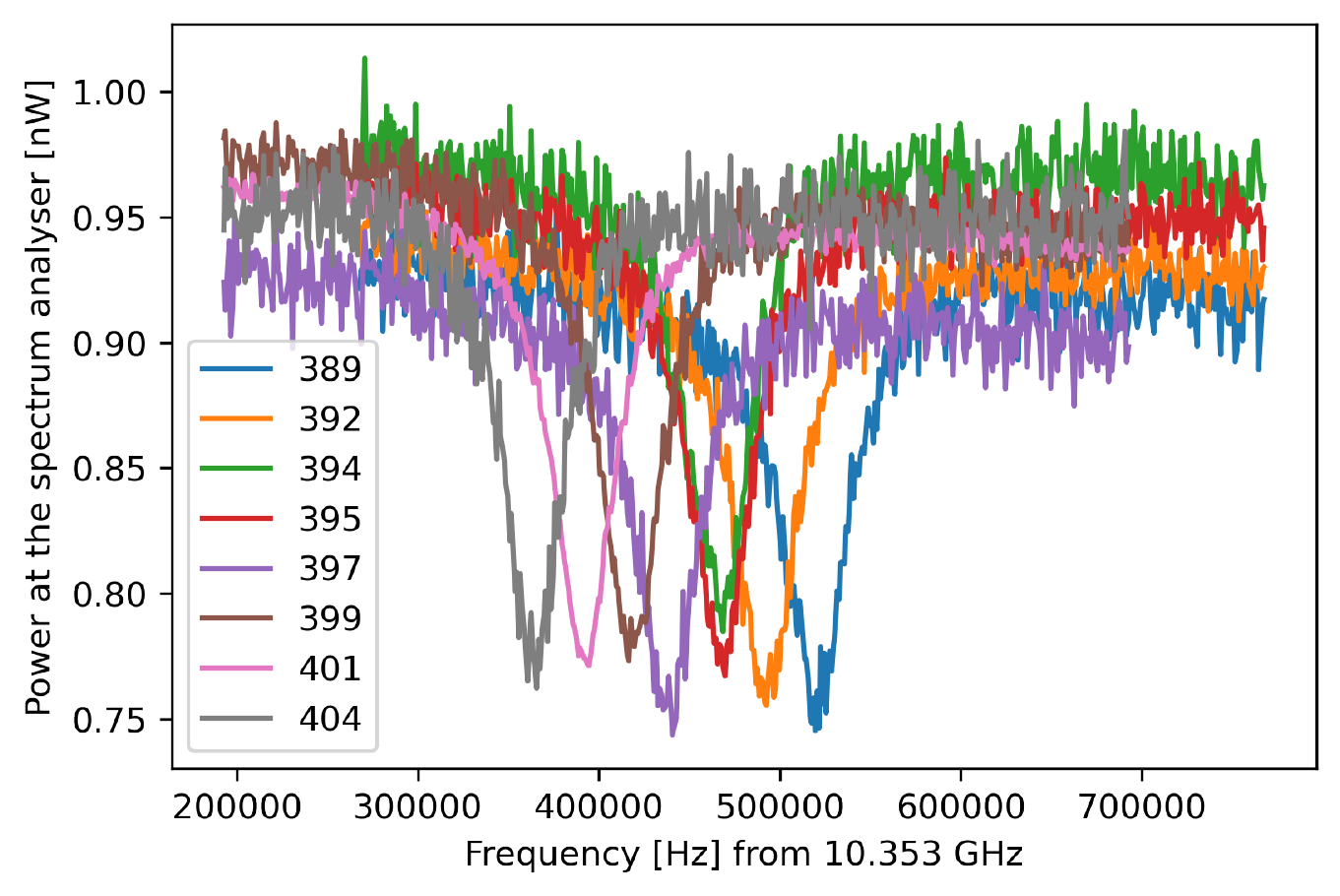}
\caption{\small Cumulative plots of the axion search measurements  with the magnet energized (action 4). Such plots are taken  using the spectrum analyser, i.e. at the point P4 of Figure \ref{fig:Apparatus2}, only for control purpose during data taking and are not used in the data analysis. The different levels outside the resonance is a measure of the gain stability of the system, kept within a few percent in a 17 hours time span. The resolution bandwidth for these plots is 1 kHz. Different colors refer to the different RUNs. }
\label{fig:vacc}
\end{figure}

Table \ref{tab:runset} summarizes all the scans performed. Figure \ref{fig:vacc} shows all the vacuum spectra measured with the spectrum analyser. Such plots are only taken for control purpose, while the data sampled by the ADC and stored in the computer are those used for the search and will be discussed in  Section \ref{dataanalysis}.

\subsubsection{Raw data processing}

The 4 s long time sequences produced by the ADC are divided into chunks about 1.5 ms long, which are Fourier transformed and averaged. Another averaging is then performed to obtain a single power spectrum ${\rm PS}_n$ for each RUN having a resolution of $B=651$ Hz and covering the down converted window $[-1$ , $+1]$ MHz. %A typical spectrum resulting from this process is shown in Figure.
A raw data processing is performed to obtain relevant parameters. This is the same procedure already described in \cite{PhysRevD.106.052007}.
In particular, the antenna coupling $\beta$ is obtained by fitting the cavity reflection spectrum measured with the spectrum analyser SA with tracking generator SG2 (see 'action 2'). The cavity loaded quality factor $Q_L$ and central frequency $\nu_c$ are obtained by fitting the average spectra obtained from the ADC data collected while the diode noise source was at the input of line L1 (see 'action 3'). The resulting parameters are reported in Table \ref{tab:runset2}. The table reports also the amplitude of the reference peak set to the frequency +900 kHz in the down converted spectra. This measurement shows again that the stability of the detection chain gain was within a few percent along the complete data taking.

\comment{

\begin{table}[htbp]
   \centering
   \begin{tabular}{|c|c|c|c|c|} % Column formatting, @{} suppresses leading/trailing space
   \hline
   RUN & $\nu_c-10.353$ GHz & Cavity $Q_L$ & $\beta$ & Ref Peak \\
   n & (Hz)  & &  & (a.u.) \\
   \hline
   389 & 522 551 & 230 000 & 10.86 & 179 \\
   \hline
   392 & 494 103 & 240 000 & 11.98 & 185\\
   \hline 
   394 & 468 841 & 245 000 & 12.17 & 186 \\
   \hline
   395 & 468 841 & 245 000 &12.17 & 187\\
   \hline
   397 & 439 835 & 245 000 & 11.43 & 175 \\
   \hline
   399 & 418 536 & 245 000 & 11.43 & 191 \\
   \hline
   401 & 393 135 & 250 000 & 11.37 & 186 \\
   \hline
   404 & 365 376 & 255 000& 11.86 & 193 \\
   \hline
   \end{tabular}
   \caption{Summary of the parameters obtained by the fits of preliminary data for all the RUNs performed. Ref Peak is the amplitude of the reference peak due to the pure tone injected by SG2 on line L3 during data taking at a frequency 900 kHz above the reference oscillator of the mixer.}
   \label{tab:runset2}
\end{table}
}

% Rounded values for frequency and corrected beta values
\begin{table}[htbp]
   \centering
   \begin{tabular}{|c|c|c|c|c|} % Column formatting, @{} suppresses leading/trailing space
   \hline
   RUN & $\nu_c-10.353$ GHz & Cavity $Q_L$ & $\beta$ & Ref Peak \\
   n & (Hz)  & &  & (a.u.) \\
   \hline
   389 & 522 600 & 230 000 & 21.6 & 179 \\
   \hline
   392 & 494 100 & 240 000 & 23.8 & 185\\
   \hline 
   394 & 468 800 & 245 000 & 24.2 & 186 \\
   \hline
   395 & 468 800 & 245 000 &24.2 & 187\\
   \hline
   397 & 439 800 & 245 000 & 22.7 & 175 \\
   \hline
   399 & 418 500 & 245 000 & 22.6 & 191 \\
   \hline
   401 & 393 100 & 250 000 & 22.5 & 186 \\
   \hline
   404 & 365 400 & 255 000& 23.5 & 193 \\
   \hline
   \end{tabular}
   \caption{Summary of the parameters obtained by the fits of preliminary data for all the RUNs performed. Ref Peak is the amplitude of the reference peak due to the pure tone injected by SG2 on line L3 during data taking at a frequency 900 kHz above the reference oscillator of the mixer. The fits errors on the values of cavity frequencies are about 100 Hz, for the loaded quality factors about 3000, and for the $\beta$'s about 0.2.}
   \label{tab:runset2}
\end{table}

During the raw analysis, a careful check of the ADC data compared with the SA data has evidenced a problem present in the down-converted data. The ADC input is filtered by a single pole low pass filter having the -3 dB point at about 1.7 MHz. Unfortunately this is not enough to avoid aliasing of the rather flat noise input. From the comparison of the high frequency and down converted spectra we estimate that the measured wide band noise is about a factor 1.7 larger with respect to  the real average noise in the vicinity of the cavity resonance. 

Each power spectrum PS$_n$ is the readout of the ADC input, and to obtain the power at the cavity output it must be divided by the overall gain. 
Alternatively, one equals the noise level measured at the ADC with the power given by the effective noise temperature of the system. We have assumed that the system noise temperature has not changed over the entire data taking time, having a duration of 17 hours. Indeed, the relative error of the $T_{\rm sys}$, about 6.3 \%,  is larger of the relative changes of the reference peak as obtained by Table \ref{tab:runset2} having a maximum of 5.1 \%. For each RUN we assume that the  noise level at the cavity frequency is 
\begin{equation}
    P_n(\nu)=k_B T_{\rm sys}^{\rm ADC} B
\end{equation}

where $B=651$ Hz is the bin width, $k_B$ is the Boltzmann's constant, and $T_{\rm sys}^{\rm ADC}= 1.7 \times T_{\rm sys} = 3.5$ K.

%\begin{figure}[htb]
%  \centering
  %    \includegraphics[width=0.45\textwidth]{RunTable.pdf}
%\caption{\small List of the scans performed.  }
%\label{fig:scan}
%\end{figure}

%
\section{\label{sec:results}Data Analysis and Results}

\label{dataanalysis}

\subsection{Axion Detection procedure}
Detection algorithms can be discussed in the classical
"hypothesis testing" framework: on the basis of the observed
data, we must decide whether to reject or fail
to reject the null hypothesis (data are consistent with noise) against
the alternative hypothesis (noise and signal are present in data) usually by means
of a detection threshold determined by the desired significance level.
%At this stage of analysis, a signal due to genuine axion-photon coupling  is indistinguishable from unmodelled noise sources due to the similar stochastic properties of noise and expected signal. 
The outcome of this data analysis step is a set of "axion candidate" masses or frequencies. However, axion signal has some distinctive properties that can be used to discriminate it from spurious detected signals (see Sect. III.B). We emphasize that the basis of our detection algorithm is  a very robust model of the noise that allow us to use maximum likelihood criterion (i.e. a $\chi^2$ test)  to implement the decision rule. Deviations from the model of the noise power spectral density are clues of excess power that could be associated with a signal. In the frequency domain, the noise model for the power spectral density at the haloscope output  (under the general assumptions of  linearity, stationarity, ergodicity and single-resonance system) is simply a first order polynomial ratio 
\begin{equation}
p(\nu)= 
\frac{\nu-\nu_z + i \gamma_z}{\nu-\nu_p + i \gamma_p}
\end{equation}
where $(\nu-\nu_{p,z} + i \gamma_{p,z})$ are the pole and the zero values in the complex plane, respectively. The fitting function to the power spectrum data, with fit parameters {${a,b,...f}$}, reads
\begin{equation}
F(\nu)= e^2*\frac{|\nu-a+ib|^2}{|\nu-c+id|^2}+f*(\nu-c)
\label{bartlet}
\end{equation}
where the linear term accounts for the slight dependence of the  ADC gain on frequency 
and we made the approximation $\nu^2-\nu^2_{p,z} \simeq 2 \nu_{p,z}(\nu-\nu_{p,z}) $. The parameter $e$ is a normalization factor.

The estimate of power spectrum resides on the Bartlett method of averaging 
periodograms \cite{Bartlett1948}. 
For every measured RUN of table \ref{tab:runset} we performed the fit with $F(\nu)$ and used the $\chi^2$ to test the hypothesis of no signal. We discovered that the quality, i.e. the value for example of the reduced $\chi^2$, of the fits was worsening with the duration of the RUN. Indeed, a key issue is the stability of our set-up, where it is actually not surprisingly that drifts will appear over time scales of several hours. For this reason we decided, for the analysis procedure, to split every RUN in subruns with a fixed length of 2000 s, and to perform the fits on each of the resulting 23 subruns. 
For each subrun, the fits are performed on a window of 200 bins (bin width = 651 Hz) centered at the cavity peak frequency. The weight on each bin is the measured value divided by $\sqrt{N}$, $N$ is the number of averages: for a 2000 s duration and 651 Hz bin width, $N\simeq1.37\cdot 10^6$.
The null hypothesis $H_0$ is accepted for $\chi^2$ probability above the chosen threshold of $P_\alpha=0.001$. Since for all subruns $H_0$ was not rejected, it is then possible to build a grand spectrum of all the residuals of the different fits, which will result in an increased sensitivity. 
\begin{figure}[htb]
  \centering
      \includegraphics[width=0.45\textwidth]{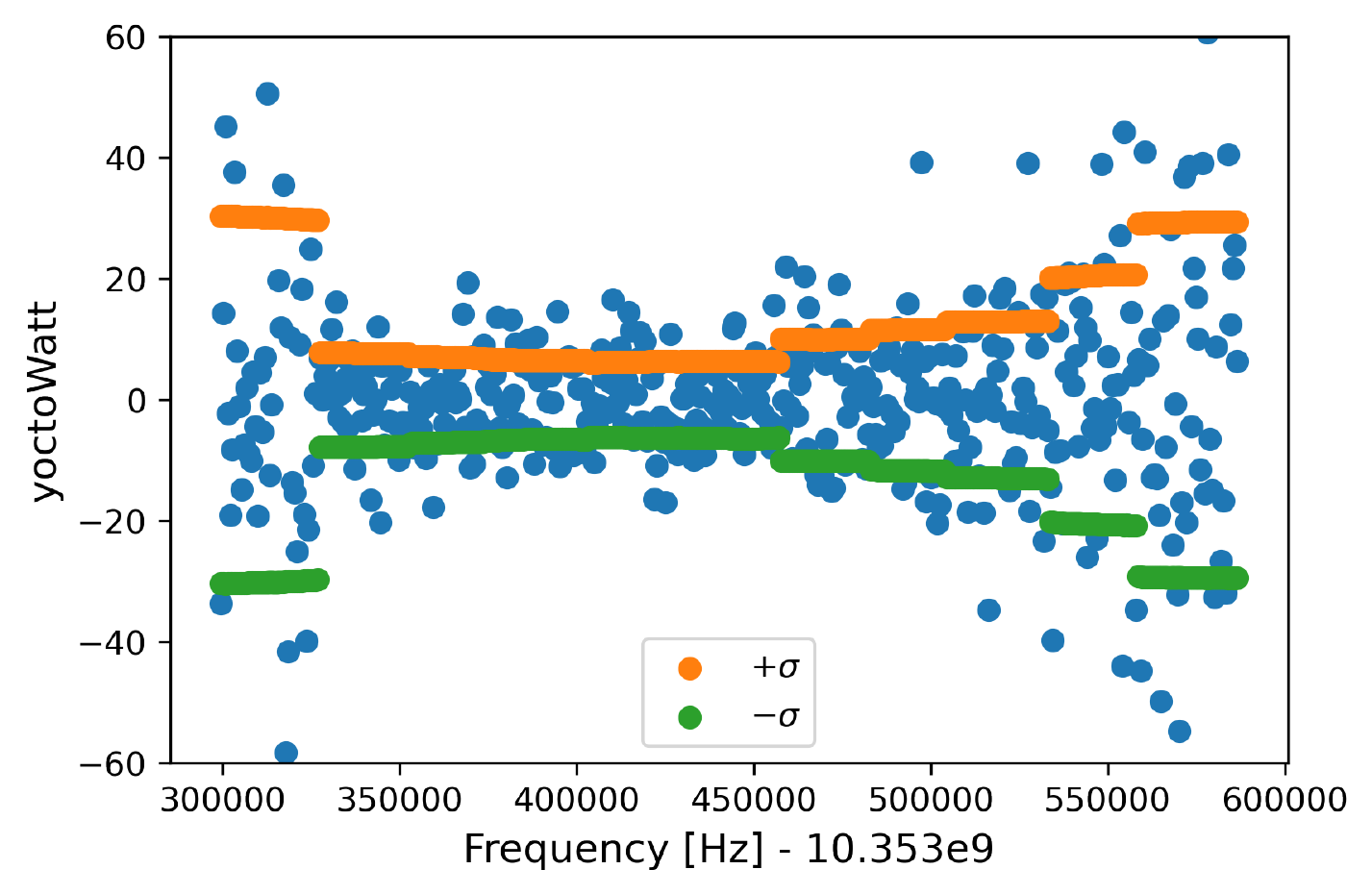}
      \includegraphics[width=0.45\textwidth]{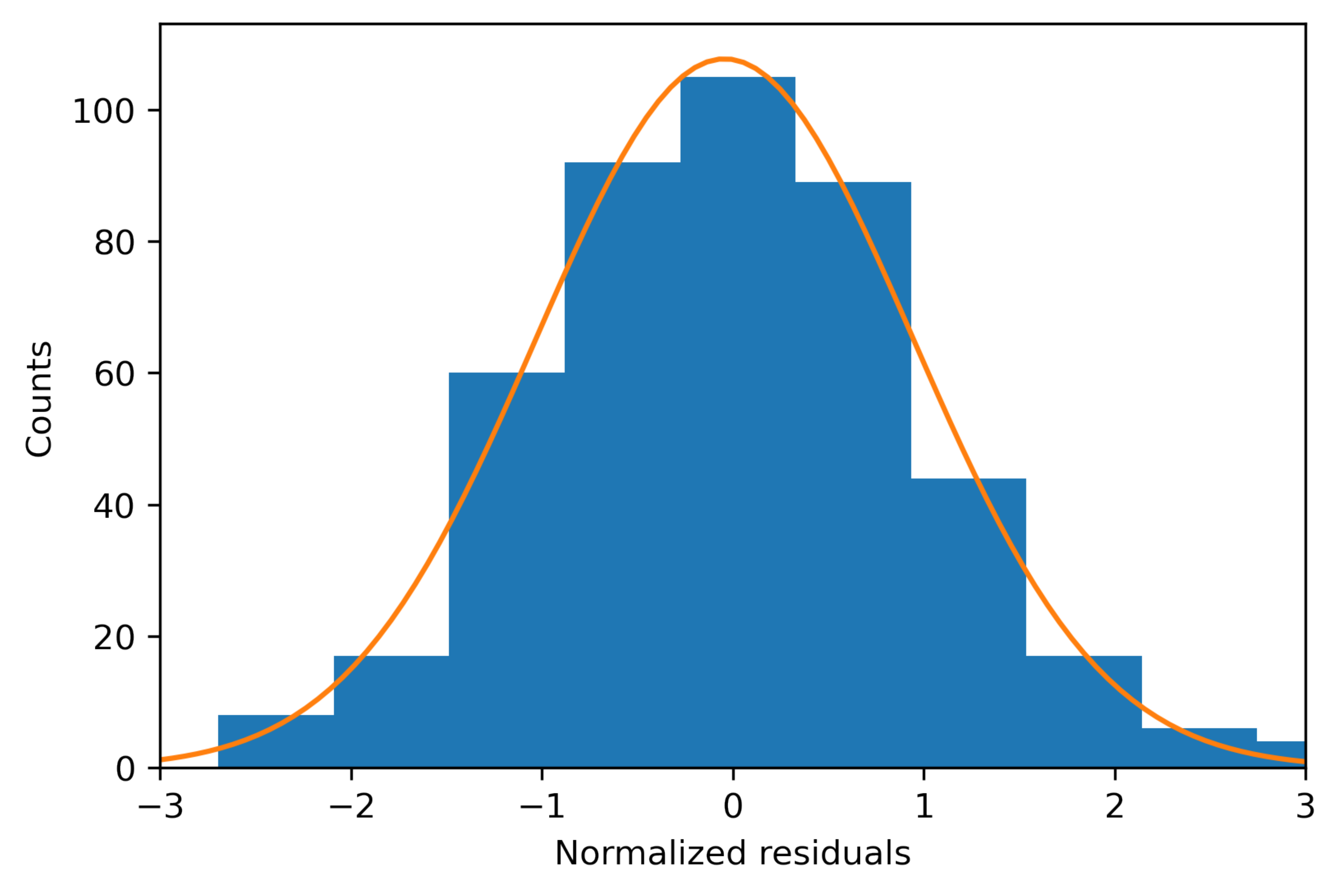}
\caption{\small Grand spectrum of the residuals. (a) Residuals and their sigma versus vs frequency. The bin size is 651 Hz. (b) Histogram of the normalized residuals. The fit function is a gaussian function with average -0.04 and width 0.99.}
\label{fig:resi}
\end{figure}

The grand spectrum is built by performing a weighted average of all the bins with the same frequency for all subruns residuals, 
%vertical weighted average of all the residual, 
again using as weight the values previously used to do the fits. 
%Vertical average means that are performed at each bin value. 
The grand spectrum of the residuals and the relative histograms of the normalized values are shown in Figure \ref{fig:resi}. The grand spectrum has 442 bins, a total $\chi^2 = 452$ having a probability $P_{\chi^2}=0.37$, and the hypothesis of data compatible with pure thermal noise is not rejected. Such a claim is done without modeling the axion.
%but imaging it is spread all over the detection bandwidth. 
%To increase the sensitivity we could assume a particular model and recalculate the $\chi^2$ over smaller intervals: this is the approach that will be used to in the calculation of the upper limits.
The minimum value of the sigma of the residuals in Figure \ref{fig:resi} is $(6.2\pm 0.4) \cdot 10^{-24}$ W, obtained at the frequency of 403645 Hz shifted from 10.353 GHz with a total integration time of 36000 s. The value expected from Dicke's radiometer equation is $(6.5 \pm 0.4)\cdot 10^{-24}$ W.

\subsection{Axion Discrimination procedures} 

Candidates that survive to a simple repetition of the $\chi^2$ test  (using a new data set taken in the same experimental conditions) can be further discriminated 
using a stationarity and on-off resonance tests. A stationarity test verifies that a signal is continuously present during a data taking. An on-off resonance test verifies that the signal can be maximized by a tuning procedure of the cavity. Moreover, the dependence of signal power on the antenna coupling  $\beta$ can also be checked. Eventually, a change of the magnetic field allow us to verify if the signal power is proportional to the magnetic field squared. Candidates that passed this step would be associated to axionic dark matter. When no axion candidate survive at the sensitivity level of current axion models\cite{di2020landscape}, an upper limit on axion-photon coupling can be set for the standard model of Galaxy halo. 
%However, such sensitivity requirements do not hold for upper limits on Axion Like Particles (ALPs) couplings. 

\subsection{Upper limits on axion-photon coupling }

Having data compatible with the presence of only noise, we can then proceed to derive bounds on the coupling constant of the axion to the photon, assuming specific coupling  and galactic halo models.
Bounds can be inferred adopting the following approach.
The power spectra for each subrun are fitted again using as fit function the sum of the background $F(\nu)$ (see Equation (\ref{bartlet})) and the expected power produced by axion conversion for specific values of the axion mass within the measured bandwidth.  
%Since we  excluded the presence of axion signals above the noise threshold,  in order to have a good fit the axion added power must be of the order of $\sigma_{\scriptscriptstyle \textup{Dicke}}$.  
By placing the axion coupling  as a free parameter, since this new fit procedure returns as output the smallest possible observable power, it is possible to obtain an upper limit on the  coupling constant $g_{a\gamma\gamma}$. Again, $\chi^2$ probability is used to evaluate the goodness of the fit.

To calculate the expected axion signal we will rely on the standard halo model for dark matter, and assume that dark matter is composed by axions in its totality. With this hypothesis, the axion energy distribution is given by a Maxwell-Boltzman distribution \cite{turner1990periodic}

\begin{equation}\label{maxwell}
f_a(\nu,\nu_a) =\frac{2}{\sqrt{\pi}}\sqrt{\nu-\nu_a}\left(\frac{3}{1.7\,\nu_a \langle\beta_a^2\rangle}\right)^{3/2} e^{-\frac{3(\nu-\nu_a)}{1.7\,\nu_a \langle\beta_a^2\rangle}}
\end{equation}

where $\nu_a$ is the axion frequency and $\langle\beta_a^2\rangle$ is the  RMS of the axion velocity distribution normalized by the speed of light. The factor 1.7 takes into account that we are working in the laboratory frame \cite{HaystacAnalysis}.  The RMS of the galactic halo is $270$  km/s, resulting in a $\langle\beta_a^2\rangle$ of $8.1\cdot 10^{-7}$.

In a haloscope, the power is released in a microwave cavity of frequency $\nu_c$ and volume $V$, immersed in a static magnetic field $B_0$. At the antenna output, with coupling $\beta$, the available power is described by the spectrum

%\begin{equation}\begin{split}
%&S_a(\nu,\nu_a,\nu_c)=\left( \frac{\hbar^3 c^3\rho_a}{m_a^2}  \right) \times \\
%&	\times \left( 2 \pi \nu_c \frac{B_0^2 V C_{030}}{\mu_0}   \right) 
 %\frac{\beta}{1+\beta} \cdot F_c(\nu,\nu_c) \cdot f_a(\nu, \nu_a) 
 %\end{split}
%\end{equation}
\begin{equation}\begin{split}
&S_a(\nu,\nu_a,\nu_c)=g_{a\gamma\gamma}^2 \frac{\hbar^3 c^3\rho_a}{m_a^2}  \times \\
&	\times \frac{2 \pi \nu_cB_0^2 V C_{030}}{\mu_0}    
 \frac{\beta}{1+\beta} \cdot F_c(\nu,\nu_c) \cdot f_a(\nu, \nu_a) 
 \end{split}
 \label{eq:power}
\end{equation}

where $\rho_a \sim  0.45$\,GeV/cm$^3$   is the local dark matter density~\cite{10.1093/ptep/ptaa104}, $g_{a\gamma\gamma}$ is the coupling constant of the axion-photon interaction, $m_{a}$ is the axion mass. The form factor $C_{030}=0.035$  has been recalculated to take into account the static magnetic field distribution over the cavity mode TM030 used in our haloscope. The function 
\begin{equation}
F_c(\nu,\nu_c)= \frac{Q_L}{1+\left[2Q_L \frac{(\nu - \nu_c)}{ \nu_c} \right]^2}
\end{equation}
describes the bandwidth limited amplification of the microwave cavity, with $Q_L$ its loaded quality factor.

\begin{figure}[htb]
  \centering
      \includegraphics[width=0.45\textwidth]{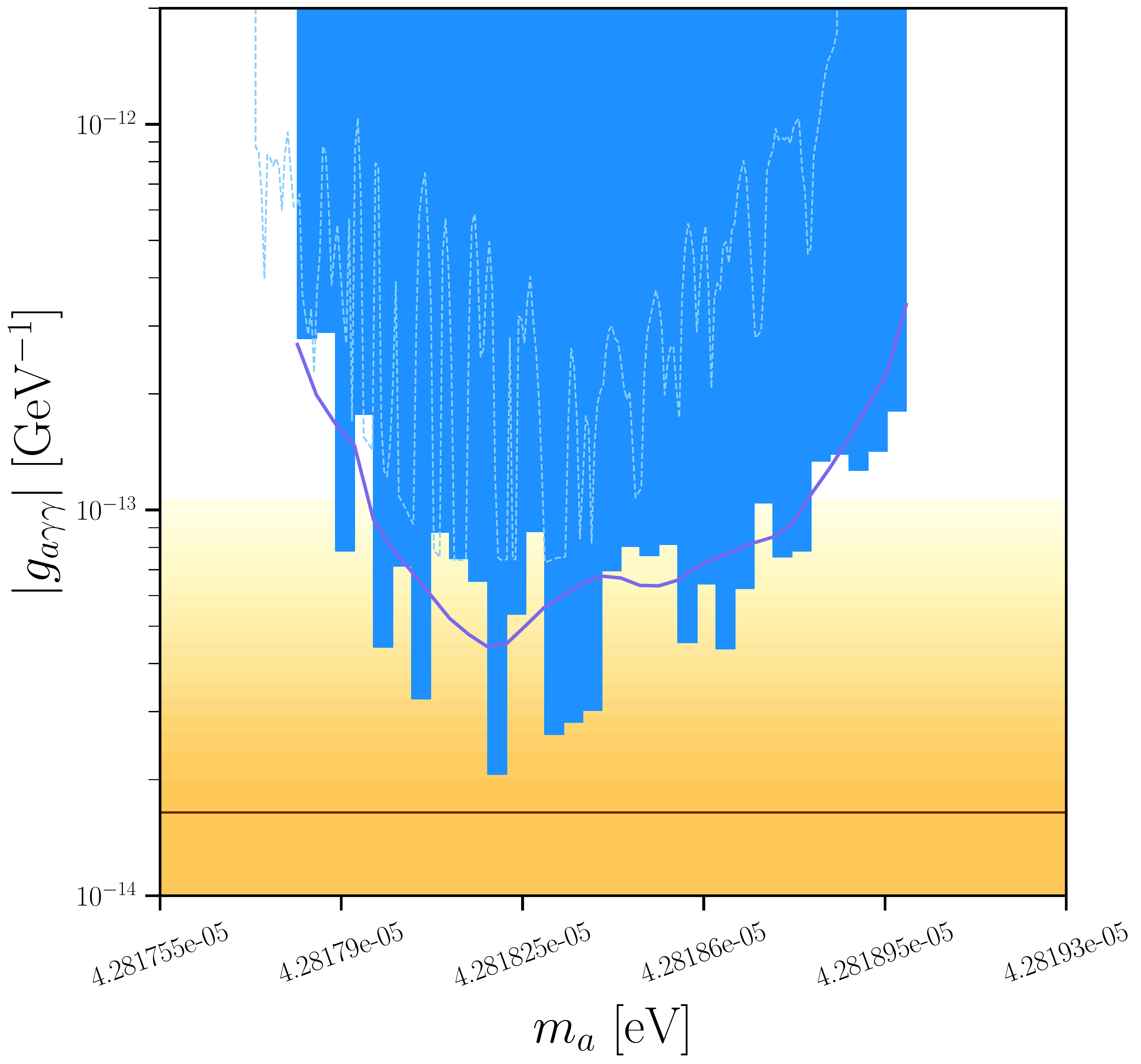}
\caption{\small $g_{a \gamma \gamma}$ axion coupling constant upper limit calculated with  90\% single-sided C.L. as a function of the axion mass. The solid blue bars are the limits obtained with the data and the analysis described in this manuscript. The solid dark blue line is the expected limit in the case of no signal. The dashed light blue line  are the limits obtained in our previous work ~\cite{PhysRevD.106.052007} . The yellow region indicates the QCD axion model band. The  horizontal line shows the theoretical prediction for the coupling constant given by the KSVZ\cite{PhysRevLett.43.103,SHIFMAN1980493} model.
%and DFSZ\cite{DINE1981199,Zhitnitsky:1980tq} models. 
Image realized with \cite{AxionLimits}.  }
\label{fig:limits}
\end{figure}
We split the measured frequency range in 33 intervals having center frequencies $\nu_i$, where we tested our sensitivity for an axion of mass $m_a= h \nu_i / c^2$ in order to obtain limits on its coupling. By using the discretized function:

\begin{equation}
 W(\nu_i,\nu_a,\nu_c)= F(\nu_i)+S_{s}(\nu_i,\nu_a,\nu_c)B
 \label{newfit}
\end{equation}

fits where performed on all the 23 subruns. 
% end comment
This fit function has $g_{a\gamma\gamma}^2$ as fitting parameter, in addition to those of $F(\nu)$ previously described.  
In order for the fits to converge, the parameters of $F(\nu)$ have as initial guess the values found when fitting only the background and they are also constrained to variate in a small interval. We have verified that such procedure is able to extract the correct value of $g_{a\gamma\gamma}^2$ by performing Monte Carlo simulations with software injected signals in our data.

%For a fixed axion mass, we fitted all the 23 subruns using equations \ref{newfit}. 
The 23 estimated set of  values  $g_{a\gamma\gamma}^2 (\nu_i)$ are then  averaged using as weights the  inverse of their squared uncertainties extracted by the fitting procedure. 
%This process is repeated for all the axion masses.
Using this approach we are able to extract a  set of $g_{a\gamma\gamma}^2$ values from which
we calculated the limit on the coupling strength with a 90\% confidence level ~\cite{alesini2021search}, adopting  a power constrained procedure for the values of $g_{a\gamma\gamma}^2$ that  fluctuates below $-\sigma_{g_{a \gamma \gamma}}$ ~\cite{cowan2011power}. 
The upper limit  $g_{a\gamma\gamma}^{\scriptscriptstyle \textup{CL}}$ obtained  adopting this procedure  in the axion mass range  $42.8178-42.8190$ $\mu$eV are reported in Figure \ref{fig:limits}  as blue bars. The minimum value sets a limit  $g_{a\gamma\gamma}^{\scriptscriptstyle \textup{CL}}<2.05\cdot 10^{-14}$~GeV$^{-1}$,  %obtained in correspondence of the maximum sensitivity (i.e.  the minimum of the  exclusion plot) 
that is 1.2 times larger respect to the benchmark KSVZ  axion models \cite{PhysRevLett.43.103,SHIFMAN1980493}.

% In presence of a signal due to axion-photon conversion, an excess power would be observable in the residuals of the power spectrum fit to the noise model. 

% %
% In the laboratory frame, the axion signal is expected to have a width of about 10 kHz ~\cite{sikivie1983experimental, turner1990periodic}. With a power spectrum with bin width $\Delta \nu=651$ Hz we expect the axion signal to be distributed over 16 consecutive bins.

% %\section{\label{sec:results}Results}

%  $\sigma_{\scriptscriptstyle \textup{Dicke}}$ calculated 
% with the Dicke radiometer equation ~\cite{dicke1946measurement}
% \begin{equation}
%   \sigma_{\scriptscriptstyle \textup{Dicke}} = k_B T_{\rm sys} \sqrt{\Delta \nu/\Delta t}\, ,
% \end{equation}
% where $T_{\rm sys}$, is the system noise-temperature,
% %of the set-up (in our case 17.2 K as described in section  2.3), 
% $\Delta \nu$ is the bin width (651 Hz) and $\Delta t$ is the integration time (3000 s).
% % The distribution of the cumulative normalized-residuals from all the datasets is shown in Fig.~\ref{fig:cumulative_residual} along with a Gaussian fit, showing a standard deviation compatible with 1.

 %
\section{\label{sec:conclusions}Conclusions}
We reported  results of the search of galactic axions using an high-Q dielectric haloscope instrumented with a detection chain based on a traveling wave parametric amplifier working close to the quantum limit.  The investigated mass range is $42.8178-42.8190$ $\mu$eV, partially already investigated by us in a previous run and not currently accessible by other experiments.
We set a limit to the axion coupling constant that is about  1.2 times larger respect to the benchmark KSVZ  axion model.
Our results demonstrate the reliability
%and versatility 
of our approach, which complements high Q factor dielectric cavities with strong magnetic field, and operation in ultra-cryogenic environment to exploit the noise performances of the TWPA based detection chain.
This is the first time a wide bandwidth quantum limited amplifier has been used in a haloscope working at high frequency, where internal losses of the components are significantly larger with respect  to frequencies in the lower octave band. A result that complements the one obtained at 5 GHz by the ADMX collaboration \cite{admxtwpa}, and paves the road
 for the exploration of the axion mass parameter space at frequencies above 10 GHz.
 
This result  improves a factor about 4 the sensitivity we obtained in our previous run in almost the same frequency range, thanks to the new amplifier and an improved description of the background in the data analysis, based on a robust noise model. 
With the implementation of anti-aliasing filters in our digitizing channels and a planned better isolation of the first stage amplifier, we expect to improve even more our sensitivity in the next run. A new type of cavity with larger effective volume and larger tuning will be put in operation for a future campaign of axion searches capable of covering a sizable range of mass values in the 40's $\mu$eV range.

\section*{Acknowledgments}

We are grateful to E. Berto, A. Benato, and M. Rebeschini for the mechanical work; F. Calaon and M. Tessaro for help with the electronics and cryogenics. We thank G. Galet and L. Castellani for the development of the magnet power supply, and M. Zago who realized the technical drawings of the system. We deeply acknowledge the Cryogenic Service of the Laboratori Nazionali di Legnaro for providing us with large quantities of liquid helium on demand.

This work is supported by INFN (QUAX experiment), by the U.S. Department of Energy, Office of Science, National Quantum Information Science Research Centers, Superconducting Quantum Materials and Systems Center (SQMS) under the Contract No. DE- AC02-07CH11359,  by the European Union's FET Open SUPERGALAX project, Grant N.863313 and by the European Union's Horizon
2020 research and innovation program under grant
agreement no. 899561. M.E. acknowledges the European
Union's Horizon 2020 research and innovation program
under the Marie Sklodowska Curie (grant agreement no.
MSCA-IF-835791). A.R. acknowledges the European
Union's Horizon 2020 research and innovation program
under the Marie Sklodowska Curie grant agreement No
754303 and the 'Investissements d'avenir' (ANR-15-
IDEX-02) programs of the French National Research
Agency.
N.C. is supported by the European Union’s Horizon 2020 research and innovation program under the Marie Skłodowska-Curie grant agreement QMET No. 101029189.

The data that support the findings of this study are available from the corresponding author upon reasonable request.

%\newpage

\appendix*

\section{Details on the experimental set up}
Table \ref{components} shows the relevant components used in the experimental set-up.

\begin{table*}[htbp]
   \centering
   \begin{tabular}{|c|c|c|c|} % Column formatting, @{} suppresses leading/trailing space
   \hline
   Components & Type & Provider/Model & Parameters @ 10 GHz \\
   \hline
   \hline
   \multicolumn{4}{|c|} {Cryogenic set up - Figure \ref{fig:Apparatus}} \\
   \hline
   K1, K2, K3 & Attenuators & Hewlett Packard 8493B 20 DB & IL = 20 dB \\
   \hline
   C1 & Circulator & Raditek RADC-8-12-Cryo-0.02-4K-S23-1WR-MS-b & IL = 0.6 dB\\
   \hline
   C2, C3 & Isolator & Raditek RADI-8-12-Cryo-0.02-4K-S23-1WR-MS-b & IL = 0.6 dB\\
   \hline
    HP & High Pass Filter & Mini Circuit VHF-7150+ & IL = 0.7 dB \\
    \hline
     Cables & RF Cable & KeyCom ULT-05 & IL =  1.9 dB/m \\
     \hline
    Cables & SC RF Cable & KeyCom NbTiNbTi085 & -- \\
     \hline
     HEMT & Amplifier&  Low Noise Factory LNF-LNC4-16B & Gain = 42 dB \\
     \hline
     \textcircled{T} & Thermometer & ICE Oxford RuO2 RCWPM 1206-68-2.21 KOHM & \\
     \hline
     $I_{\rm DC}$ & Current source & Keithley 263 & \\
     \hline
     \hline
        \hline
   \multicolumn{4}{|c|} {Room temperature set up - Figure \ref{fig:Apparatus2}} \\
        \hline
   K4 & Attenuator & Narda Micro-Pad 4779 - 12 & IL = 12 dB \\
    \hline
  Cables   & RF Cable & Huber - Suhner SF104 & IL = 1 dB/m \\
     \hline
     CD1 & Power Splitter & Macom 1147 & \\
     \hline
     CD2 & Power Combiner & Triangle Microwave YL - 74 & \\
\hline
     HEMT & Amplifier & Low Noise Factory LNF-LNR4-16B & Gain = 35 dB\\
     \hline
     AI, AQ & Amplifier & Femto DPHVA-101 & Gain (1 MHz) = 50 dB \\
          \hline
     Mixer & Mixer & Miteq IRM0812LC2Q & -- \\
     \hline
     A1 & Amplifier & MITEQ AFS4-08001200-10-CR-4 & Gain = 32 dB\\
     \hline
     Noise Source& Noise Source & Micronetics NS346B  & $T_{\rm eff} = 10 000$ K \\
     \hline
     SG1, SG2, LO, PUMP & Signal Generator & Keysight N5183/N5173 & \\
     \hline
     SA & Spectrum Analyser & Keysight N9010B & \\
     \hline
     ADC & Analog to digital Converter & National Instruments USB 6366 & Rate = 2 Ms/s\\
     \hline
     \hline
         \hline
   \multicolumn{4}{|c|} {Other components} \\
     \hline
     Dewar &  & Precision Cryogenics System Inc. & \\
     \hline
     Dilution Unit & & Leiden Cryogenics &\\
     \hline
     Magnet & & Cryogenic Ltd & \\
     \hline
     
   \end{tabular}
   \caption{Description of the relevant components for the experimental setup. IL = Insertion Loss.}
   \label{components}
\end{table*}

%\newpage

%\end{acknowledgments}

%\input{manualBiblio.tex} %uncomment only if you want manual bibliography
\bibliography{RunTWPA}

% If in two-column mode, this environment will change to single-column
% format so that long equations can be displayed. Use
%\begin{widetext}
%\end{widetext}

\end{document}